\documentclass[12pt,letterpaper]{article}
\pdfoutput=1
\usepackage{graphicx,array}
\usepackage{color}
\usepackage{latexsym}
\usepackage{amsthm}
\usepackage{amsmath}
\usepackage{amssymb}
\usepackage{hyperref}
\usepackage{bbold}

\def\simgt{\mathrel{\lower2.5pt\vbox{\lineskip=0pt\baselineskip=0pt
           \hbox{$>$}\hbox{$\sim$}}}}
\def\simlt{\mathrel{\lower2.5pt\vbox{\lineskip=0pt\baselineskip=0pt
           \hbox{$<$}\hbox{$\sim$}}}}

\newcommand{\be}{\begin{equation}}
\newcommand{\ee}{\end{equation}}
\newcommand{\bea}{\begin{eqnarray}}
\newcommand{\eea}{\end{eqnarray}}
\newcommand{\Eq}[1]{Eq.~(\ref{#1})}

\newcommand{\MM}{\mathcal{M}}
\newcommand{\NN}{\mathcal{N}}

\newcommand{\sslash}[2]{\ensuremath\raisebox{0.03cm}{{\slash}}\hspace{#2cm}#1\/}

\renewcommand{\l}{\langle}
\renewcommand{\r}{\rangle}

\newcommand{\pdet}{\textrm{pdet}}
\newcommand{\GLn}{GL(n,\mathbb{C})}

\definecolor{nicered}{rgb}{0.7,0.1,0.1}
\definecolor{nicegreen}{rgb}{0.1,0.5,0.1}
\hypersetup{colorlinks,citecolor= nicegreen,linkcolor= nicered}

\begin{document}
\hfill

\vspace{4cm}

\begin{center}
{\LARGE\bf
 Gravity Amplitudes from $n$-Space
}\\
\bigskip\vspace{1cm}{
{\large Clifford Cheung$^{1,2}$}
} \\[7mm]
 {$^1$\it Department of Physics, University of California,
          Berkeley, CA 94720} \\
 {$^2$\it Theoretical Physics Group, Lawrence Berkeley National Laboratory,
          Berkeley, CA 94720}
 \end{center}
\bigskip
\centerline{\large\bf Abstract}

\begin{quote} \small

We identify a hidden $GL(n,\mathbb{C})$ symmetry of the tree level $n$-point MHV gravity amplitude.  Representations of this symmetry reside in an auxiliary $n$-space whose indices are external particle labels.
Spinor helicity variables transform non-linearly under $GL(n,\mathbb{C})$, but linearly under its notable subgroups, the little group and the permutation group $S_n$.  Using $GL(n,\mathbb{C})$ covariant variables, we present a new and simple formula for the MHV amplitude which can be derived solely from geometric constraints.
This expression carries a huge intrinsic redundancy which can be parameterized by a pair of reference 3-planes in $n$-space.  Fixing this redundancy in a particular way, we reproduce the $S_{n-3}$ symmetric form of the MHV amplitude of \cite{Hodges2}, which is in turn equivalent to the $S_{n-2}$ symmetric form of \cite{Wen} as a consequence of the matrix tree theorem.  The redundancy of the amplitude can also be fixed in a way that fully preserves $S_n$, yielding new and manifestly $S_n$ symmetric forms of the MHV amplitude.  Remarkably, these expressions need not be manifestly homogenous in spinorial weight or mass dimension.  We comment on possible extensions to N$^{k-2}$MHV amplitudes and speculate on the deeper origins of $GL(n,\mathbb{C})$.

\end{quote}

\newpage

\section{Geometry of $n$-Space}

In this paper we argue that the tree level $n$-point MHV gravity amplitude possesses a hidden $\GLn$ symmetry.   This symmetry acts on an auxiliary $n$-space indexed by the labels of external particles, $a \in \{1,2,\ldots,n\}$.  We will show that geometric constraints in $n$-space are sufficient to derive a new and simple expression for the MHV gravity amplitude in terms of $\GLn$ covariant objects.  By virtue of $\GLn$ this formula is manifestly symmetric under the permutation group $S_n$.
Our formula contains a large redundancy which originates from the geometric constraints.  Fixing this redundancy yields the expressions for the MHV gravity amplitude in the literature, as well as new ones.
 
Consider a general $n$-point amplitude, written as a function of spinor helicity variables, $\lambda_{a\alpha}$  and $\tilde \lambda^{a\dot\alpha}$.  What is the action of $\GLn$ on these kinematic variables?
Naively, it is  natural to define the spinors  $\lambda_{a\alpha}$  and $\tilde \lambda^{a\dot\alpha}$ as 2-planes which transform as fundamental and anti-fundamental representations of $GL(n,\mathbb{C})$, so
\bea
\lambda_{a\alpha} \rightarrow\sum_b  \lambda_{b\alpha} G_{\;\;a}^{b} \quad &,&\quad
\tilde \lambda^{a\dot\alpha} \rightarrow \sum_b \tilde \lambda^{b\dot\alpha} G^{-1a}_{b} .
\eea
Indeed, for this choice the condition of momentum conservation 
\bea
\sum_a \tilde \lambda^{a\dot\alpha} \lambda_{a\alpha} &=&0,
\eea
is manifestly $\GLn$ symmetric and implies that these 2-planes are mutually orthogonal.  
As such, $\lambda_a$ and $\tilde \lambda^a$ provide a natural linear representation of $\GLn$.  

However, just because $\GLn$ acts on the particle label $a$  does not imply that  $\GLn$ must act linearly on $\lambda_a$ and $\tilde \lambda^a$.  These representations can, in principle, be
 non-linear functions of $\lambda_a$ and $\tilde \lambda^a$.  
In the present work we argue that for MHV gravity amplitudes the correct variables are instead
\bea
v_a^i &=&\frac{i}{2}\sum_{\alpha,\beta} \lambda_{a\alpha} \sigma^{i\alpha\beta} \lambda_{a \beta} \quad ,\quad {i\in\{1,2,3\}} \nonumber \\
\phi^{ab} &=&\frac{[ab]}{\l ab\r} \quad , \quad  a\neq b.
\label{eq:definitions}
\eea
Here $v_a^i$ is a symmetric spinor product transforming as a $(1,0)$ representation of the Lorentz group, and the diagonal elements $\phi^{aa}$ are as of yet undefined.  The variable  $\phi^{ab}$ was defined in the recent remarkable papers of Hodges on the MHV amplitude \cite{Hodges2,Hodges1}.
Our claim is that $v_a^i$ and $\phi^{ab}$ furnish linear representations of $\GLn$, so
\bea
v_{a}^i &\rightarrow& \sum_b v_{b}^i G_{\;\;a}^{b} \nonumber \\
\phi^{ab} &\rightarrow& \sum_{c,d}\phi^{cd} G^{-1a}_{c} G^{-1b}_{d}. 
\label{eq:GLntransform}
\eea  
Note the placement of raised and lowered indices---under $\GLn$, the 3-plane $v_a^i$ transforms as a fundamental and the tensor $\phi^{ab}$ transforms as a symmetric anti-bi-fundamental.  Moreover, \Eq{eq:definitions} and \Eq{eq:GLntransform} imply that $\lambda_a$ and $\tilde \lambda^a$ transform non-linearly under $\GLn$.

  From $v_a^i$ we can now construct the Lorentz invariant, $\GLn$ covariant tensors
\bea
v_{ab} &=& \sum_i v_a^i v_b^i =\frac{1}{2} \l a b\r^2 \nonumber \\
v_{abc} &=& \sum_{i,j,k}\epsilon^{ijk} v_a^i v_b^j v_c^k =\frac{1}{2} \l a b\r\l b c\r\l c a\r.
\label{eq:vabc}
\eea
Obviously, additional tensors can be constructed from higher order products, but only the quantities above will be important for the MHV gravity amplitude.

Lastly, let us consider the little group and the permutation group $S_n$, which are subgroups of $\GLn$.    A little group transformation sends $\lambda_a \rightarrow w_a \lambda_a$ and $\tilde \lambda^a \rightarrow w_a^{-1}\tilde\lambda^a$, corresponding to the diagonal subgroup of $\GLn$ defined by
\bea
 G^b_{\;\;a}    &=&\delta^b_{\;\;a} w_a^2,
 \label{eq:weight}
 \eea
in \Eq{eq:GLntransform}.
Likewise, the action of $S_n$ is $\lambda_a \rightarrow \lambda_{\sigma(a)}$ and $\tilde \lambda^a \rightarrow \tilde \lambda^{\sigma(a)}$, corresponding to
\bea
 G^b_{\;\;a}    &=&\delta^b_{\;\;\sigma(a)}.
 \label{eq:sigma}
 \eea
Thus, even though the $\GLn$ covariant variables $\phi^{ab}$ and $v_a^i$ are non-linear functions of $\lambda_a$ and $\tilde \lambda^a$, this is consistent with the fact that these  spinors must transform linearly under the little group and $S_n$.

\section{Amplitudes from Geometry}

We assume that the MHV gravity amplitude is a function of $\phi^{ab}$ and $v_a$ alone, so in terms of on-shell  $\NN=8$ superspace, the amplitude is
\bea
\MM^{\rm MHV}_n &=& \delta^{2 \NN}\left(\sum_a  \lambda_{a} \tilde \eta_a\right) \hat \MM^{\rm MHV}_n(\phi^{ab},v_a).\eea
Furthermore, we assume the crucial projection relation,
\bea
\sum_b \phi^{ab} v_b^i &=&0,
\label{eq:proj}
\eea
whose importance was rightly emphasized in \cite{Hodges2}.
As we will see, the underlying $\GLn$ symmetry together with the geometric relation in \Eq{eq:proj} entirely fix the structure of the MHV amplitude!

It is sometimes convenient to re-express the projection condition in component form
\bea
\sum_b \phi^{ab} |b\r | b\r =0.
\label{eq:linrel}
\eea
By dotting the left hand side with a symmetric product of reference spinors, $\l x|\l y| + \l y | \l x|$, we can solve for the diagonal components $\phi^{aa}$ yielding  
\bea
\label{eq:offdiag}
\phi^{aa} &=& -\sum_{b\neq a} \frac{[ab] \l b x \r \l b y\r}{\l ab\r \l ax \r \l a y\r},
\eea
precisely as defined in \cite{Hodges2}.  Of course, the explicit dependence on $|x\r$ and $|y\r$ always cancels, since these spinors are only here to provide a component form expression of the geometric constraint in \Eq{eq:proj}.

What is $\hat \MM^{\rm MHV}_n$?  Certainly, the amplitude should depend on $\phi^{ab}$ and $v_a$ in such a way that leaves no free $a$ indices---otherwise the MHV amplitude would not preserve $\GLn$.  The first obvious choice is $\sum_{a,b,i} \phi^{ab} v_a^i v_b^i$, but this vanishes due to the projection condition in \Eq{eq:proj}.
Another natural possibility is $\det (\phi^{ab})$, but this also vanishes because \Eq{eq:proj} implies that $\textrm{rank}(\phi^{ab})=n-3$.  Nonetheless, there is an alternative invariant quantity which is suitable: the pseudo-determinant, defined as the product of all non-zero eigenvalues.  For a general matrix $M$, the pseudo-determinant is formally defined as
\bea
\pdet (M) &\equiv& \lim_{\epsilon \rightarrow 0} \epsilon^{-\textrm{null(M)}}\det(M + \epsilon \mathbb{1}),
\eea
where null$(M)$ is the nullity of $M$.  
Unfortunately, $\pdet(\phi^{ab})$ yields an expression which does not have uniform little group weight, since $\phi^{ab}$ is covariant while the identity matrix is invariant.  To get around this we construct a vierbein defined by the geometric relations
\bea
\sum_a  v_a^i  e^{a}_{\;\;\alpha}&=& \delta^{i}_{\;\;\alpha}, \quad \alpha \in\{1,2,\ldots n\}, \quad i \in\{1,2,3\} \nonumber \\
\sum_a e^{\alpha}_{\;\;a} e^{a}_{\;\;\beta} &=& \delta^{\alpha}_{\;\;\beta}.
\label{eq:vierbein}
\eea
The constraint in \Eq{eq:vierbein} obviously does not fix the vierbein $e^{a}_{\;\; \alpha}$ or inverse vierbein $e^{\alpha}_{\;\; a}$ uniquely, since $v_a^i$ is a 3-plane.    Specifically, the definition in \Eq{eq:vierbein} carries an intrinsic redundancy parameterized by the $(n-3)$-plane orthogonal to $v_a^i$.
This redundancy in the vierbein will later manifest itself in the MHV gravity amplitude.  

Introducing ``left'' and ``right'' vierbeins which separately satisfy \Eq{eq:vierbein}, we can transform $\phi^{a b}$ into an orthonormal basis,
\bea
\phi^{ab} &\rightarrow&  \sum_{a,b} \phi^{ab} e_{La}^\alpha e_{Rb}^\beta , 
\label{eq:phialphabeta}
\eea
whose $\pdet$ can be taken.  The natural invariant that can be constructed in the orthonormal basis is
\bea
 \det(e_{L\alpha}^{a})\det(e_{R\alpha}^{a}) \pdet\left(\sum_{a,b} \phi^{ab} e_{La}^\alpha e_{Rb}^\beta\right),
\label{eq:step1}
\eea
where the $\det$ prefactors have been included so that this object has the little group weight of an MHV gravity amplitude.
Note that unlike $\det$, $\pdet$ does not, in general, distribute multiplicatively over products, so the vierbeins inside and outside of the $\pdet$ cannot be canceled against each other.

The expression in \Eq{eq:step1} can be recast in slightly more palatable form by reshuffling the vierbein fields and rewriting the $\pdet$.  In particular,
\bea
 \pdet\left(\sum_{a,b} \phi^{ab} e_{La}^\alpha e_{Rb}^\beta\right) &=& \lim_{\epsilon\rightarrow 0} \epsilon^{-3}\det\left(\sum_{a,b} \phi^{ab} e_{La}^\alpha e_{Rb}^\beta + \epsilon \mathbb{1}^{\alpha \beta} \right) \nonumber \\
 &=& \det\left(\sum_{a,b} \phi^{ab} e_{La}^\alpha e_{Rb}^\beta +  \sum_i \delta^{\alpha i}\delta^{\beta i} \right) \nonumber \\
 &=&  \det(e_{L\alpha}^{a})^{-1}\det(e_{R\alpha}^{a})^{-1} \det\left( \phi^{ab} +  \sum_{\alpha, \beta, i} e^a_{L\alpha}e^b_{R\beta} \delta^{\alpha i}\delta^{\beta i} \right).
\eea
Note that in going from the first to the second lines, it was  crucial that $\phi^{ab}$ satisfy the projection relation in \Eq{eq:proj} and the vierbeins satisfy the defining relation in \Eq{eq:vierbein}.   Given these geometric constraints, the second line follows from the first line as a matter of linear algebra---it has nothing to do with the kinematic information information implicitly contained in the auxiliary variables.

Combining the above expression with \Eq{eq:step1} yields our final result: a new form of the MHV gravity amplitude which is $\GLn$ symmetric:
\begin{center}
\framebox{
\parbox[t][3.0cm]{12cm}{

\addvspace{0.4cm} \centering 
\bea
\hat \MM^{\rm MHV}_n &=& \frac{1}{4} \det\left( \phi^{ab} +  \sum_{\alpha, \beta, i} e^a_{L\alpha}e^b_{R\beta} \delta^{\alpha i}\delta^{\beta i} \right).\qquad
\label{eq:wow1}
\eea
} 
}\end{center}
Here the constant normalization of the amplitude is not fixed by $\GLn$, so we take it from the known expression. 
We emphasize that ${\it any}$ choice of the vierbeins $e_{L\alpha}^{a}$ and $e_{R\alpha}^{a}$ which satisfies \Eq{eq:vierbein} will produce the correct amplitude.  However, as noted earlier,  \Eq{eq:vierbein} carries a huge ambiguity in the definition of $e_{L\alpha}^{a}$ and $e_{R\alpha}^{a}$.  This redundancy can be parameterized as a pair of $(n-3)$-planes. In an abuse of nomenclature, we will refer to this redundancy as a ``gauge freedom'' which can be ``gauge fixed'' by an appropriate choice of reference $(n-3)$-planes.  As we will see, gauge fixing the redundancy reproduces the known MHV formulas in the literature.

 Under the $\GLn$ transformation in \Eq{eq:GLntransform}, the MHV amplitude transforms as
\bea
\hat \MM^{\rm MHV}_n &\rightarrow & \det(G^a_{\;\;b})^{-2} \hat \MM^{\rm MHV}_n.
\label{eq:density}
\eea
Restricting $G^a_{\;\;b}$ to little group transformations, we see that \Eq{eq:density} implies that \Eq{eq:wow1} has the proper spinorial weight.
The transformation law in \Eq{eq:density} strongly suggests that the MHV amplitude is secretly a density or integral measure over $\GLn$ covariant variables.

\Eq{eq:wow1} can be expressed in an alternative but equivalent form which is more simple for computations.  
In particular, note that 
\bea 
\sum_{\alpha, \beta,i} e^a_{L\alpha}e^b_{R\beta} \delta^{\alpha i}\delta^{\beta i} &=& \sum_i e^{ai}_{L}e^{bi}_{R} ,
\eea is simply the outer product of a pair of 3-planes, which is a rank 3 matrix.  However, this is not any old rank 3 matrix, since it is constrained by  the definition of the vierbeins in \Eq{eq:vierbein}.  Nonetheless, \Eq{eq:wow1} can be reassembled in terms of an arbitrary rank 3 matrix,  at the cost of some Jacobian factors which depend on $v_a^i$.  Without loss of generality, we can parameterize this arbitrary rank 3 matrix with $\sum_i L^{ai} R^{ai}$, where  $L^{ai}$ and $R^{ai}$ are arbitrary 3-planes.   This effectively dualizes the pair of $(n-3)$-planes which parameterize the gauge freedom into a pair of 3-planes. 
Using linear algebraic identities, we find that \Eq{eq:wow1} has a second equivalent form,
\begin{center}
\framebox{
\parbox[t][4.0cm]{12cm}{

\addvspace{0.4cm} \centering 
\bea
\hat\MM^{\rm MHV}_n =  \frac{\det\left(\phi^{ab}+ \sum\limits_iL^{ai}R^{bi}\right)}{\left(\sum\limits_{a,b,c}L^{abc} v_{abc}/3\right)\left(\sum\limits_{d,e,f}R^{def}v_{def}/3\right)},\qquad
\label{eq:myformula}
\eea
} 
}\end{center}
where the denominators are simple Jacobian factors, and in analogy with \Eq{eq:vabc} we have defined
\bea
L^{abc} = \sum_{i,j,k} \epsilon^{ijk} L^{ai}L^{bj}L^{ck} \nonumber \\
R^{abc} = \sum_{i,j,k} \epsilon^{ijk} R^{ai}R^{bj}R^{ck}.
\eea
\Eq{eq:myformula} holds for {\it any} choice of the reference 3-planes, $L^{ai}$ and $R^{ai}$.
Note that while \Eq{eq:myformula} is manifestly $S_n$ symmetric, it is {\rm not} simply a summation over permutations of an existing form of the MHV gravity amplitude.

\section{Equivalent Representations from Gauge Fixing}

Our expression for the MHV amplitude in \Eq{eq:myformula} has a huge redundancy parameterized by the reference 3-planes, $L^{ai}$ and $R^{ai}$.  In this section we show how a very particular gauge fixing of this redundancy yields the formula of Hodges \cite{Hodges2}.  We then prove that the formulas of \cite{Hodges2} and \cite{Wen} are equivalent as a consequence of the matrix tree theorem.
Afterwards, we consider reference 3-planes which are fully $S_n$ symmetric.  These gauge choices yield new, manifestly $S_n$ symmetric forms of the MHV amplitude.  
\subsection{Old Representations}

The very simplest gauge fixing of $L^{ai}$ and $R^{ai}$ yields the formula of \cite{Hodges2}.  In particular, we set
\bea
L^{ai} &=& \delta^{a l_i} \nonumber\\
R^{ai} &=& \delta^{a r_i},
\label{eq:gf}
\eea
where $l_i$ and $r_i$ each denote an ordered triplet of external particles.  Thus, $\sum_i L^{ai} R^{ai}$ is an $n\times n$ matrix whose entries are all zero except at $\{ l_i, r_i\}$ for $i=1,2,3$.  \Eq{eq:gf} implies that 
\bea
\det\left(\phi^{ab} + \sum_i L^{ai} R^{ai}\right) &=& \det\left(\phi^{ab}_{l_1l_2l_3,r_1r_2r_3}\right),
\eea
where $\phi^{ab}_{l_1l_2l_3,r_1r_2r_3}$ is the $(n-3)\times(n-3)$ reduced matrix obtained by removing the rows $l_i$ and the columns $r_i$.  For simplicity we ignore the overall sign of the amplitude, since it is unimportant and anyway just changes by $\textrm{sgn}(\{l_1 l_2 l_3 12 \ldots \sslash{l_1}{-.16}\sslash{l_2}{-.16}\sslash{l_3}{-.16} \ldots n\} \rightarrow \{r_1 r_2 r_3 12 \ldots \sslash{r_1}{-.23}\sslash{r_2}{-.23}\sslash{r_3}{-.23} \ldots n \})$.
Finally, plugging \Eq{eq:gf} into \Eq{eq:myformula} yields
\bea
\hat \MM_n^{\rm MHV} &=& \frac{ \det(\phi^{ab}_{l_1l_2l_3,r_1r_2r_3})}{\l l_1 l_2 \r \l l_2 l_3 \r \l l_3l_1 \r\l r_1 r_2 \r \l r_2 r_3 \r \l r_3 r_1 \r},
\label{eq:Hodges}
\eea
which is precisely the Hodges form of the MHV gravity amplitude.
  As proven in \cite{Hodges2}, any choice for $l_i$ and $r_i$ and \Eq{eq:Hodges} will produce the same final answer.   The maximally permutation symmetric choice is of course $l_i=r_i$, which preserves a manifest $S_{n-3}$ symmetry.

The formula of \cite{Hodges2} can be easily shown to be equivalent to the manifestly $S_{n-2}$ symmetric expression of \cite{Wen}.  To show this, we rescale the entries of $\phi^{ab}$ so that the full matrix has uniform weight,
\bea
\tilde \phi^{ab} &=& \phi^{ab} \times \l ax\r \l ay \r \l bx\r \l by \r,
\label{eq:rescale}
\eea
which implies that
\bea
\tilde \phi^{aa} &=& -\sum_{b\neq a} \tilde \phi^{ab} ,
\label{eq:sum}
\eea
so the elements of any row or column of $\tilde \phi^{ab}$ sum to zero.  Note that this relation is an algebraic identity fixed by \Eq{eq:proj}.

Plugging in $\{x,y\}=\{1,2\}$ for the reference spinors and setting $l_i = r_i=\{1,2,3\}$, we find that \Eq{eq:Hodges} becomes
\bea
\hat \MM_n^{\rm MHV} 
&=&\frac{\det( \phi_{123,123}^{ab})}{\l 12 \r^2 \l23 \r^2 \l 31\r^2} \nonumber\\
&=& \frac{1}{\l 12\r^2}  \left( \prod_{a>2}^n \frac{1}{\l a 1\r^2 \l a 2\r^2} \right)  \det(\tilde \phi_{123,123}^{ab}),
\eea
where the products of angle brackets arise from the rescaling of the rows and columns of $\phi^{ab}$ to go to $\tilde \phi^{ab}$ variables.

From \Eq{eq:rescale} it is clear from our choice of reference spinors that the first and second rows and columns of $\tilde \phi^{ab}$ vanish identically.  Thus we need only consider the reduced $(n-2)\times(n-2)$ matrix, $\tilde \phi_{12,12}^{ab}$.   By definition, $\det(\tilde \phi_{123,123}^{ab})$ is a cofactor of the matrix $\tilde \phi_{12,12}^{ab}$ with the third row and column removed.
Since $\tilde \phi_{12,12}^{ab}$ is by construction $S_{n-2}$ symmetric on its remaining indices, this implies that all cofactors are equal, so
\bea 
\label{eq:pdetpre}
\det(\tilde \phi_{123,123}^{ab}) &=& \det(\tilde \phi_{12l_3,12l_3}^{ab}),
\eea
 for $l_3 \in \{3,\ldots, n\}$.
Any matrix whose rows and columns sum to zero possesses a null eigenvector whose components are all equal.  This eigenvector is invariant under $S_n$ transformations.  As is well-known, for such a matrix all cofactors are the same and equal to
\bea
 \det(\tilde \phi_{12l_3,12l_3}^{ab})&=& (n-2)^{-1}\; \pdet(\tilde \phi_{12,12}^{ab}),
\label{eq:pdet}
\eea 
where $\pdet$ is the product of all eigenvalues modulo for the zero corresponding to the $S_n$ invariant null eigenvector.
The $(n-2)$ symmetry factor in \Eq{eq:pdet} arises from the fact that the index $a$ in \Eq{eq:pdetpre} can take on that many values.

The indices of $\tilde \phi_{12,12}^{ab}$ correspond to the external points $\{3,4,\ldots,n\}$.  As in \cite{Wen}, we can identify these legs with the vertices of a graph, and associate with each pair $\{a,b
\}$ a corresponding link.  With this mapping, \Eq{eq:pdet} is recognized as a formulation of the matrix tree theorem, which says that 
\bea
(\textrm{\# of  trees})&=& (\textrm{\# of vertices})^{-1} \; \pdet (\textrm{Laplacian matrix}),
\label{eq:matrixtree}
\eea
where the Laplacian matrix is defined as the difference between the degree matrix and the adjacency matrix.
Like $\tilde \phi^{ab}$,  the Laplacian matrix has rows and columns which sum to zero.
Hence, \Eq{eq:pdet} is a ``weighted'' version of \Eq{eq:matrixtree} in which each link of the tree is accompanied by the corresponding factor from $\tilde\phi^{ab}$.
 This yields the expression
\bea
\hat \MM_n^{\rm MHV} &=& \frac{1}{\l 12\r^2}  \left( \prod_{a>2}^n \frac{1}{\l a 1\r^2 \l a 2\r^2} \right) \sum_{\rm trees} \prod_{\textrm{edges } ab} 
\frac{[ab]}{\l ab \r} \times \l a1\r \l a2 \r \l b1\r \l b2 \r,
\eea
where the summation is over all possible trees which span a labeled graph with $(n-2)$ vertices.
This proves the equivalence of the MHV formulae in \cite{Hodges2} and \cite{Wen}.

\subsection{New Representations}

One can go beyond the formulas in the existing literature by opting for more exotic gauge fixings.  Most gauge fixings of $L^{ai}$ and $R^{ai}$ will explicitly break the $S_n$ symmetry of the amplitude, since these are 3-planes residing in an $n$-dimensional space.  However, there is a $S_n$ covariant 3-plane which can be chosen as the reference, namely
\bea
L^{ai}  = R^{ai} = v_{a}^i.
\eea
 Note, crucially, the difference in the raised and lowered operators on the left and right hand sides. While this gauge fixing is $S_n$ covariant, it explicitly breaks the full $\GLn$.  However, because this $\GLn$ breaking enters precisely through a gauge redundancy, the final answer is still $\GLn$ invariant, as required by \Eq{eq:wow1}.

With this gauge fixing, we obtain a new, {manifestly} $S_n$ symmetric form of the MHV gravity amplitude
\bea
\hat \MM^{\rm MHV}_n &=& \left(\sum_{a,b,c} v_{abc}^2/3\right)^{-2}\det\left(\phi^{ab}+v_{ab}\right) \nonumber \\
&=& \left(\sum_{a,b,c}\frac{\l ab\r^2 \l bc \r^2 \l ca\r^2}{12}\right)^{-2}\det\left(\phi^{ab}+\frac{\l ab\r^{2}}{2}\right)
\label{eq:perm1}
\eea
Let us comment on some remarkable features of this formula.  As noted earlier, this gauge fixing explicitly breaks $\GLn$.  Since the little group is a subgroup of the broken $\GLn$, this expression is not manifestly homogenous under spinorial weights.  This is clear, for example, from the denominator factor, which is a sum over monomials of different spinor weight.  Furthermore, the expression is not manifestly homogenous in mass dimension since $\phi^{ab}$ is a dimensionless phase while $v_{ab}$ has mass dimension 2.  Of course, \Eq{eq:perm1} still gives the final correct answer, which  has both correct spinorial weight and correct mass dimension.
This was guaranteed by the underlying gauge redundancy in the MHV amplitude.

There is another $S_n$ symmetric choice of reference 3-planes which can be made using the anti-holomorphic spinor $\tilde \lambda^a$.  In particular, we choose
\bea
L^{ai}  = R^{ai} &=& \tilde v^{ai}
\eea
where we have defined the anti-holomorphic 3-plane in the obvious way,
\bea
\tilde v^{ai} &=&
\frac{i}{2}\sum_{\alpha,\beta} \tilde \lambda^{a\dot\alpha} \bar\sigma^{i}_{\;\dot \alpha \dot \beta} \tilde \lambda^{a \dot\beta} .
\eea
This yields yet another $S_n$ symmetric form of the amplitude,
\bea
\hat \MM^{\rm MHV}_n 
&=& \left(\sum_{a,b,c}\frac{s_{ab} s_{bc} s_{ca}}{12}\right)^{-2}\det\left(\phi^{ab}+\frac{[ ab]^{2}}{2}\right).
\eea
While the above expression is clearly little group covariant, it is inhomogeneous in mass dimensions.  Again, this does not matter because of the intrinsic gauge redundancy of the MHV amplitude.

\section{Future Directions}

This paper leaves numerous possible directions for future work.  The leading open question is conjectural: might $\GLn$ be a hidden symmetry of all gravity amplitudes? To evaluate this possibility, an understanding of the space of geometric constraints relevant to the tree level N$^{k-2}$MHV amplitudes will be essential.  Given the substantial evidence for a hidden $\GLn$ at the MHV level, it may even be that higher loop amplitudes can be similarly constructed.

The key players in this paper were the auxiliary variables $\phi^{ab}$ and $v_a^i$.  We have seen hints that these quantities are, fundamentally, integration variables which have been localized to their values in \Eq{eq:definitions}.  This is certainly suggested by the transformation law in \Eq{eq:density}, which would arise from an integration measure over $\GLn$ covariant auxiliary variables.   In this way, gravity amplitudes could nicely mirror the Grassmannian twistor formulation of $\NN=4$ SYM \cite{ACCK}, which has been discussed in great depth in numerous papers \cite{twistor, ACCK,DCI,loop}, among others.
There is also a likely connection between our results and the recently proposed formulas for $\NN=8$ gravity \cite{Cachazo1,Cachazo2}, which have already been studied in some recent work \cite{He2,Bullimore,Cachazo:2012pz}.

\vspace*{0.5cm}

\noindent{\it Note added: During the completion of this work \cite{He} and \cite{Mason} also pointed out the equivalence of the MHV amplitudes in \cite{Hodges2} and \cite{Wen} using the matrix tree theorem.} 

\section*{Acknowledgements}

C.~C.~is supported by the Director, Office of Science, Office of High Energy and Nuclear Physics, of the US Department of Energy under Contract DE-AC02-05CH11231, and by the National Science Foundation under grant PHY-0855653.  C.C.~is indebted to Nima Arkani-Hamed for a timely reminder of C.C's earlier unpublished note from 2009 relating MHV amplitudes to the matrix tree theorem.



\begin{thebibliography}{}
  
\bibitem{Hodges2}
  A.~Hodges,
  arXiv:1204.1930 [hep-th].

\bibitem{Wen}
  D.~Nguyen, M.~Spradlin, A.~Volovich and C.~Wen,
  JHEP {\bf 1007}, 045 (2010)
  [arXiv:0907.2276 [hep-th]].

\bibitem{Hodges1}
  A.~Hodges,
  arXiv:1108.2227 [hep-th].

\bibitem{He} 
  B.~Feng and S.~He,
  arXiv:1207.3220 [hep-th].

\bibitem{Mason} 
  T.~Adamo and L.~Mason,
  arXiv:1207.3602 [hep-th].
  
\bibitem{twistor} 
  N.~Arkani-Hamed, F.~Cachazo, C.~Cheung and J.~Kaplan,
  JHEP {\bf 1003}, 110 (2010)
  [arXiv:0903.2110 [hep-th]].

\bibitem{ACCK} 
  N.~Arkani-Hamed, F.~Cachazo, C.~Cheung and J.~Kaplan,
  JHEP {\bf 1003}, 020 (2010)
  [arXiv:0907.5418 [hep-th]].

\bibitem{DCI} 
  N.~Arkani-Hamed, F.~Cachazo and C.~Cheung,
  JHEP {\bf 1003}, 036 (2010)
  [arXiv:0909.0483 [hep-th]].
  
\bibitem{loop} 
  N.~Arkani-Hamed, J.~L.~Bourjaily, F.~Cachazo, S.~Caron-Huot and J.~Trnka,
  JHEP {\bf 1101}, 041 (2011)
  [arXiv:1008.2958 [hep-th]].
  
\bibitem{Cachazo1} 
  F.~Cachazo and Y.~Geyer,
  arXiv:1206.6511 [hep-th].
  
\bibitem{Cachazo2} 
  F.~Cachazo and D.~Skinner,
  arXiv:1207.0741 [hep-th].
  
\bibitem{He2} 
 S.~He,
  arXiv:1207.4064 [hep-th].
  
  \bibitem{Bullimore} 
 M.~Bullimore,
  arXiv:1207.3940 [hep-th].

\bibitem{Cachazo:2012pz} 
  F.~Cachazo, L.~Mason and D.~Skinner,
  arXiv:1207.4712 [hep-th].


  
\end{thebibliography}
\end{document}